# Photometry and Spectroscopy of the Symbiotic Binary V1413 Aquilae during the 2024 Eclipse

**David Boyd**
*West Challow Observatory, West Challow, Oxfordshire, UK; davidboyd@orion.me.uk*

**David Cejudo**
*El Gallinero Observatory, Madrid, Spain; davcejudo@hotmail.com*

**James Foster**
*Pinon Pines RoR Observatory, Frazier Park, CA; jrfcomet@hotmail.com*

**Forrest Sims**
*Desert Celestial Observatory, Apache Junction, AZ; forrest@simsaa.com*

**Gary Walker**
*Maria Mitchell Observatory, Nantucket, MA; bailyhill14@gmail.com*

**Abstract**   We report our photometric and spectroscopic observations and analysis of the 2024 eclipse of the symbiotic binary V1413 Aquilae. We found the system in a visually bright state and the eclipse time of minimum consistent with the published ephemeris. The eclipse profile showed that the hot component was an extended object rather than an isolated white dwarf. By analyzing the eclipse profile we estimated the orbital inclination to be $67.9°$, the radius of the extended hot component surrounding the white dwarf to be $39.3\,R_\odot$, and that the red giant star was probably filling its Roche Lobe. From our flux calibrated spectra, we determined the brightest component of the system to be the hot component whose continuum and emission lines together are responsible for 83% of the V-band light. The circumbinary nebula and its emission lines contribute over 14%, while the red giant is responsible for less than 3%. Our spectra revealed a rich harvest of low ionization emission lines. By measuring how flux in these emission lines varied through the eclipse, we have provided information which should prove useful for future modelling of this symbiotic system.

## 1. Introduction

Symbiotic stars are binary systems with long orbital periods comprising a cool red giant star and a small hot luminous object, usually a white dwarf, which orbits in the tenuous outer atmosphere of the red giant. Hydrogen-rich material transfers from the giant to the white dwarf either by Roche-lobe overflow or by accretion from the stellar wind of the giant and accumulates on the surface of the white dwarf. Mass transfer by Roche-lobe overflow may lead to the formation of an accretion disc around the white dwarf. Orbital periods typically range from hundreds to thousands of days. The spectral energy distribution of symbiotic stars generally comprises a continuum from the cool giant star, radiation peaking in the UV from the white dwarf, and emission from a circumbinary nebula ionized by radiation from the white dwarf.

Munari (2019) describes two types of symbiotic stars, burning and accreting-only. In burning symbiotics, material accumulating in a non-degenerate form on the surface of the white dwarf undergoes steady thermonuclear burning which maintains it at a high temperature. Radiation from the hot white dwarf generates emission lines in the spectrum. Changes in the rates of mass transfer and nuclear burning may cause outbursts in which the system brightens by several magnitudes before slowly returning to a quiescent state. In accreting-only symbiotics, the nuclear material accumulates quietly on the white dwarf and emission lines are absent from the spectrum. Increasing pressure within the accreted layer may eventually trigger a symbiotic nova explosion in which the accreted layer plus material from the white dwarf are violently ejected and the star brightens by eight or more magnitudes before slowly fading and the process of accreting hydrogen-rich material begins again. In time this process will lead to another nova explosion. If the white dwarf is massive and/or the accretion rate is high, pressure may build up quickly enough on the white dwarf that more than one nova explosion on the star has been recorded. It is then called a recurrent nova. For further information on symbiotic stars in general see Mikołajewska (2010) and Munari (2019) and references therein.

## 2. Current knowledge of V1413 Aql

V1413 Aql (= AS 338) was included as entry 338 in a table of stars with strong Hα emission published by Merrill and Burwell (1950). Its first mention as a symbiotic star was in a catalogue of symbiotic stars published by Allen (1984). As shown in the AAVSO visual light curve (Kloppenborg 2024; Figure 1), it has experienced outbursts of about 3 magnitudes, reaching visual magnitude 11 on four occasions since 1989, the most recent in 2020, indicating that V1413 Aql is a nuclear burning symbiotic. After each outburst it slowly faded until the decline was interrupted by the next outburst.

Outbursts of V1413 Aql occur when there is an increase in thermonuclear activity causing a rapid expansion of the photosphere of the hot component. This causes a small increase in luminosity but because of the expansion, the temperature of the hot component decreases. It emits less in the UV and more in the visual region of the continuum so it appears to brighten. As the temperature of the hot component is lower, only lines with low ionization potential are produced in the nebula. When the activity dies down, the hot component shrinks, its temperature rises, and emission increases in the UV and decreases in the visual. The appearance of lines with high ionization potential such as He II 4686 in the nebula is an indication that the star has reached a high temperature state. The low brightness phase is sometimes referred to as quiescence (Esipov *et al.* 2000). V1413 Aql has experienced low brightness states several times in the last 35 years, the most recent being in 2017, when it faded close to magnitude 14 and He II 4686 emission briefly appeared (Tatarnikova *et al.* 2020). Its recent behavior indicates V1413 Aql is accreting material from the giant star at a sufficiently high rate that it remains in an active state and frequent outbursts are triggered. Prior to the eclipse discussed in this paper, V1413 Aql was at V magnitude 12 and still in a lower temperature, visually bright state.

Reports on previous observations of V1413 Aql and its eclipses have been published by Munari (1992), Esipov *et al.* (2000, 2013), Kolotilov *et al.* (2001), Siviero *et al.* (2007), and Tatarnikova *et al.* (2009, 2018, 2020).

V1413 Aql has deep eclipses which occur every 434.1 days as the hot component is eclipsed by the red giant. The eclipse minimum ephemeris published in Munari (1992) is:

$$T(min) = 2446650\,(\pm 15) + 434.1\,(\pm 0.2) \times E. \qquad (1)$$

The spectral type of the red giant in V1413 Aql has been reported as M5III (Munari 1992), M3 (Murset and Schmid 1999), M4 (Bełczyński *et al.* 2000), and M4.5III (Gałan *et al.* 2023). For our analysis we will adopt M5III as this appears to be the most frequently cited value in the literature. Drawing on various sources, Gałan *et al.* (2023) give its effective temperature as ~3400 K.

According to Schlafly and Finkbeiner (2011) total galactic reddening in the direction of V1413 Aql is E (B–V) = 0.60 ± 0.01. Esipov *et al.* (2000) give the color excess of V1413 Aql as E (B–V) = 0.50 ± 0.05. Gaia DR3 (Bailer-Jones *et al.* 2021) gives the distance of V1413 Aql as 5.6 (−1.0 +1.4) kpc. At that distance the star will have experienced a large proportion of the total galactic reddening. Adopting E (B–V) = 0.50 ± 0.05 and the canonical value R (V) = 3.1 for the ratio of total visual extinction A (V) to selective extinction E (B–V), we get the total extinction of V1413 Aql in the V band, A (V) = 1.55 ± 0.15.

## 3. Observatories and equipment

We operate five privately-funded observatories located on two continents, which have been used to observe the 2024 eclipse of V1413 Aql.

*West Challow Observatory* (DB) is in South Oxfordshire, UK, at an elevation of 84 m. Photometric data are acquired with a 0.35-m Schmidt-Cassegrain Telescope (SCT) operating at f/6.3 equipped with Astrodon B and V filters and an SXVR-H9 CCD camera, and are analyzed with AIP4WIN (Berry and Burnell 2005). Spectroscopic data are obtained with a 0.28-m SCT at f/10, Shelyak LISA spectroscope, spectral resolving power 1000, with a 23-μm slit and SXVR-H694 CCD camera. Spectra are analyzed using ISIS spectral analysis software (Buil 2021).

*El Gallinero Observatory* (DC) is located in Spain about 50 km north of Madrid at an elevation of 1054 m. It contains a 0.35- m Ritchey-Chretien scope at f/8.2 equipped with a LISA spectroscope, spectral resolving power 1000, with 35-μm slit and ATIK 460EX CCD camera. Photometric data are obtained with a ZWO ASI FF65 refractor, Baader B and V filters, and ZWO ASI290 CMOS camera. Photometry is analyzed with AIP4WIN and spectroscopy with ISIS.

*Pinon Pines Estate Observatory* (JF) is located 12 km west of Frazier Park in southern California at an elevation of 1686 m. For spectroscopy it is equipped with a 0.33-m Cassegrain at f/7.5, Shelyak ALPY600, spectral resolving power 600, with an ATIK414EX CCD camera and LISA IR spectroscope, spectral resolving power 1000, with an ATIK460EX CDD camera. Photometric data are obtained with a 0.35-m SCT at f/7 with Astrodon B, V, and I$_c$ filters and a SBIG ST-8xme CCD camera, and are analyzed with MAXIM DL (Diffraction Limited 2024). Spectroscopic data are analyzed with ISIS.

*Desert Celestial Observatory* (FS) is located at Apache Junction, 54 km east of downtown Phoenix, Arizona, at an elevation of 637 m. Spectroscopic data are obtained with a 0.5-m PlaneWave CDK scope, LISA spectroscope, spectral resolving power 1000, with 23-μm slit and ATIK414EX CCD camera, photometry with an Explore Scientific AR102 refractor with Astrodon B and V filters and a ZWO ASI2600MM CMOS camera. Photometric data are analyzed with AstroImageJ (Collins *et al.* 2017) and spectroscopy with ISIS.

*First Light Observatory* (GW) is located at Stars End Observatory, near Mayhill, New Mexico, at an elevation of about 7000 ft. Photometric data are obtained with a 0.6-m PlaneWave CDK scope, FLI Kepler KL400 CMOS camera, Astrodon B and V filters, and are analyzed with MAXIM DL.

Analysis of our data made extensive use of custom Python coded development using Astropy (Astropy Collab. *et al.* 2018) and also the spectral analysis package PlotSpectra (PlotSpectra 2024).

## 4. Analysis of photometric observations

Analysis of our filtered photometric observations followed normal procedures of bias, dark, and flat correction and differential aperture photometry using comparison stars from the AAVSO charts (AAVSO VSP 2024) for V1413 Aql with either AIP4WIN, MAXIM DL, or AstroImageJ. The filters used conformed to the Johnson-Cousins standard. Each observer chose comparison stars appropriate to their field of view. Times were recorded as Julian Date. Figure 2 shows our B, V, and I$_c$ magnitude measurements. The mean uncertainties in B and V were 0.011 and in I$_c$ 0.012. Ingress to the eclipse was not well covered but egress was well observed in three filters with a distinct change in slope towards the end of the eclipse. Our B and V magnitude measurements were uploaded to, and are available from, the AAVSO International Database (AAVSO AID 2024) and/or the BAA photometry database (Br. Astron. Assoc. Photometry Database 2024a).

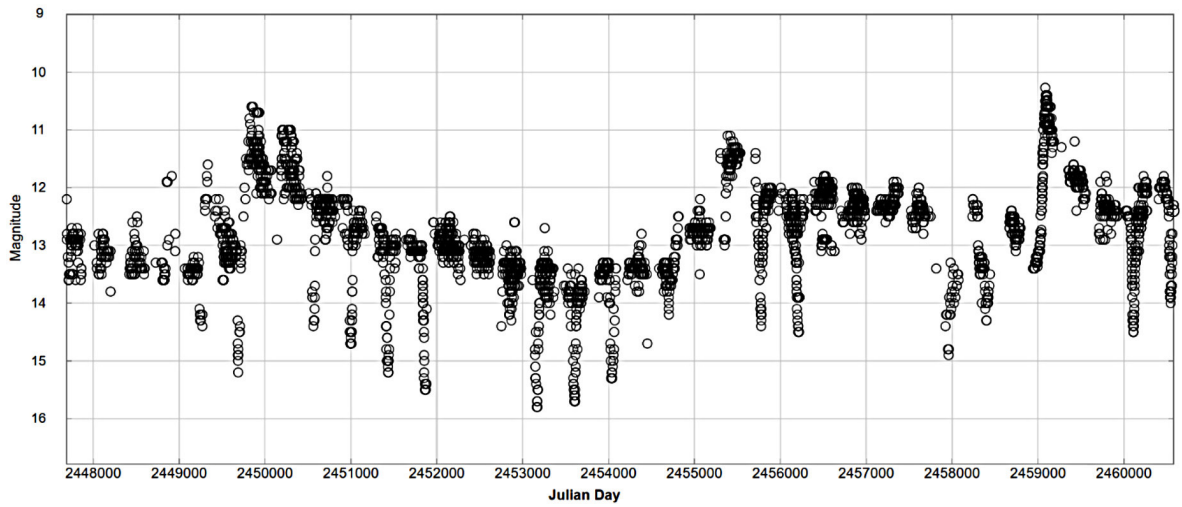

Figure 1. AAVSO visual light curve of V1413 Aql from 1989 to 2024.

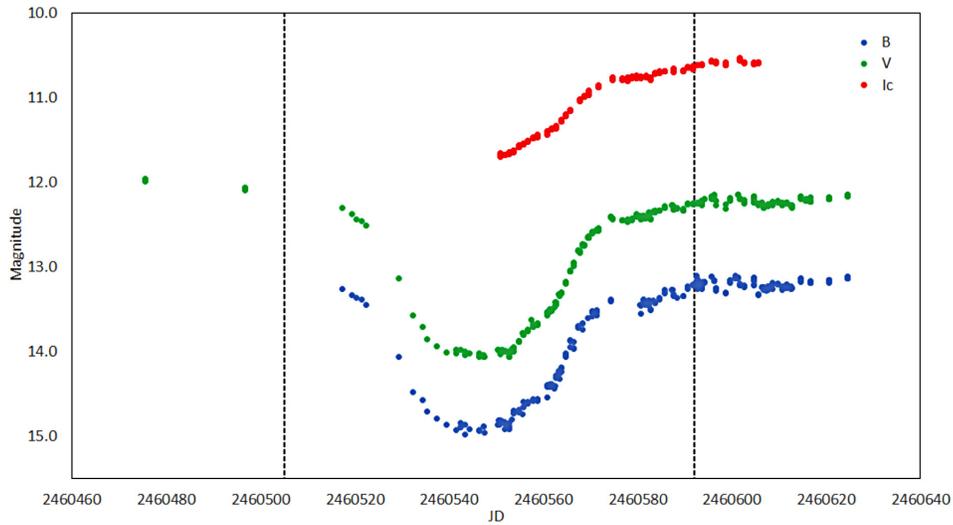

Figure 2. B, V, and I$_c$ magnitudes of V1413 Aql during the 2024 eclipse. Start and end of the eclipse are marked with vertical lines.

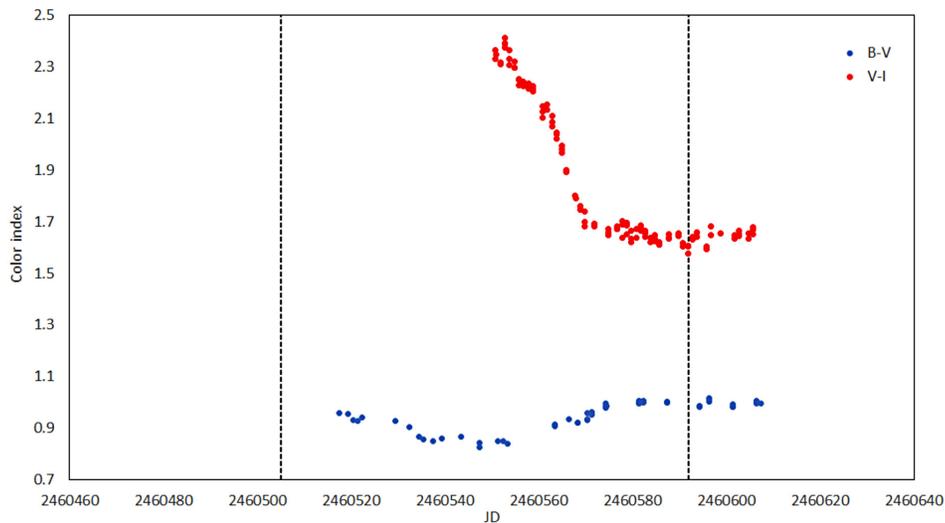

Figure 3. B–V and V–I$_c$ color indices of V1413 Aql during the 2024 eclipse. Start and end of the eclipse are marked with vertical lines.

We obtained closely concurrent B and V magnitudes before, during, and after the eclipse which were used to calculate B–V color indices. During the eclipse minimum the mean B–V color index was $0.85 \pm 0.02$ and outside eclipse the mean B–V was $1.00 \pm 0.02$. We also obtained concurrent V and $I_c$ magnitudes during and after the eclipse. During the eclipse minimum the mean V–$I_c$ color index was $2.35 \pm 0.03$ while outside the eclipse the mean V–$I_c$ was $1.64 \pm 0.03$. Our B–V and V–$I_c$ color indices are shown in Figure 3. The increase in V–$I_c$ during the eclipse is because the M5 red giant contributes a significantly greater proportion of the light of the system when the hot component is hidden. V–$I_c$ for an M5 giant is 3.19 (Worthey 1994).

According to Esipov *et al.* (2000) and Tatarnikova *et al.* (2018), a criterion for V1413 Aql being in an active state is that B–V has a value in the range $0.9 \pm 0.2$. Together with the absence of lines of high ionization potential, our observed value of B–V indicates V1413 Aql was still in an active state at the time of the eclipse. The decrease in B–V during the eclipse suggests that the part of the circumbinary nebula which remains visible during the eclipse is bluer, and therefore presumably hotter, than the eclipsed hot component including the white dwarf. With $E(B–V) = 0.50 \pm 0.05$, the intrinsic B–V color index of V1413 Aql outside eclipse is $(B–V)_0 = 0.5 \pm 0.05$.

From Schlafly and Finkbeiner (2011), with mean galactic interstellar extinction $R(V) = 3.1$, the ratio $E(V–I_c)/E(B–V) = 1.24$, although other values may be found in the literature. Assuming this value, $E(V–I_c)$ for V1413 Aql is $0.62 \pm 0.06$ and the intrinsic V–$I_c$ color index of V1413 Aql outside eclipse is $(V–I_c)_0 = 1.02 \pm 0.07$.

It has been suggested, for example in Esipov *et al.* (2013), that the hot component might contain an accretion disc. Flickering is often associated with the process of accretion and manifests as stochastically distributed overlapping flares that, in cataclysmic variables, occur on time-scales of the order of minutes (Bruch 2021). Variability in V1413 Aql reminiscent of flickering on timescales of hours and days has been reported by Merc *et al.* (2024) in TESS data, although they acknowledge the possibility of contamination by the nearby star. As flickering is stronger at shorter wavelengths, we examined our longest continuous B band runs of 30-sec cadence photometry (~2 hours) for evidence of flickering but failed to find any.

## 5. Analysis of spectroscopic observations

Spectroscopic images were analyzed using ISIS spectral analysis software (Buil 2021). This involved bias, dark, and flat correction, geometrical correction, sky background subtraction, spectrum extraction, and wavelength calibration using the integrated ArNe source in the LISA and ALPY slit spectroscopes. Spectra were then corrected for instrumental and atmospheric losses using spectra of a nearby reference star with a known spectral profile from the MILES library of stellar spectra (Falcón-Barroso *et al.* 2011) situated as close as possible in airmass to the target star and obtained immediately prior to the target spectra. Different reference stars were chosen by each observer according to their location and observing time to satisfy the close airmass criterion. Our spectra were uploaded to, and are available from, the ARAS database (ARAS Spectral Database, ARAS Group 2024) and/or BAA spectroscopy database (Br. Astron. Soc. Spectroscopy Database 2024b).

A problem in slit spectroscopy where the star remains fixed in the centre of the slit is wavelength-dependent losses due to atmospheric dispersion because of the limited width of the list. With our equipment and observing locations, atmospheric instability and tracking and guiding imperfections during multi-minute integrations introduce random motion of the stellar image within the slit which averages out the effect of wavelength-dependent slit losses. The use of reference star spectra, which suffer the same losses as spectra of the target, to correct for instrumental and atmospheric losses also helps to mitigate the problem.

Both photometry and spectroscopy were complicated by the proximity of a 13th magnitude star 12 arcsec NE of V1413 Aql. Care was taken to choose photometry apertures which excluded this star. When V1413 Aql became fainter than this star, and on-slit guiding on V1413 Aql itself became unreliable, spectroscopy was guided on another brighter field star.

Where possible, spectra were calibrated in absolute flux by the observer using the method described in Boyd (2020). This involves measuring the V magnitude of the star at the same time as recording its spectrum. For about one third of our spectra no concurrently measured V magnitude was available. In order to estimate the V magnitude of V1413 Aql at the time of each of these spectra, the eclipse was divided into four regions covering ingress, minimum, egress rise, and exit from eclipse, and our V magnitudes in each region were fitted with a 4th order polynomial. These polynomials were then used to interpolate a V magnitude to the time of each spectrum for calibrating it in absolute flux. The RMS residual between these polynomial fits and our magnitude measurements was 0.026 magnitude. We used formulae in Cardelli *et al.* (1989) to calculate $A(\lambda)/A(V)$ and applied this wavelength-dependent dereddening to each of our flux calibrated spectra.

We found the six strongest emission lines of V1413 Aql in our spectra were redshifted from their rest wavelengths by a mean of 115 km/s. After applying a correction to account for Earth's motion around the solar system barycentre, and assuming these lines originate in the circumbinary nebula rather than on one of the components, we obtain a mean systemic radial velocity of the binary system of 92 km/s. At time of writing Gaia DR3 does not give a radial velocity for V1413 Aql.

## 6. Eclipse time of minimum

To determine the time of minimum we used multiple iterations of the Kwee-van Woerden and polynomial fit algorithms in Peranso (Paunzen and Vanmunster 2016) to analyze the lower section of the eclipse profile. These gave the mean time of minimum as JD 2460545.8 ± 0.3 (2024 August 23).

This is 4.6 days later than predicted by the Munari (1992) ephemeris but within its published uncertainty.

The eclipse in June 2023 was observed visually and photometrically as reported in Poyner (2023). Poyner gave the time of minimum as JD 2460111.5, 4.4 days later than predicted by the Munari ephemeris and consistent with our result.

Figure 4 shows data from the AAVSO AID for the 2023 eclipse in CV and V magnitudes and the 2024 eclipse in V magnitude plotted relative to orbital phase of the 2024 eclipse. The profiles are very similar.

## 7. Eclipse analysis

Figure 5 shows V-band magnitudes of V1413 Aql from the AAVSO AID in early 2024 prior to the eclipse together with our V magnitudes, all in black. There is a steady downward trend which continued after the eclipse at a mean rate of 1.88 mmag/day. This is marked as a black dotted line and the multiple polynomials fitted to our V magnitudes as described earlier are marked as a red dotted line. Our magnitudes measured during the eclipse incorporated this downward trend as the star continued its slow path back towards quiescence. We removed this downward trend before analyzing the profile of the eclipse by interpolating its trajectory over the eclipse and subtracting this from our V band measurements in such a way that the V magnitude at the midpoint of the eclipse was unchanged.

Figure 6 shows this detrended eclipse profile relative to orbital phase with phase zero at our measured time of minimum. The upper horizontal dotted line is the detrended trajectory of the out-of-eclipse V magnitude interpolated above the eclipse. Further dotted lines mark detrended magnitudes of key stages in our analysis of the eclipse. Where necessary, the points marking the ends of these lines were interpolated between our measurements.

Total duration of the eclipse is the time between the two points labelled A where light starts to slowly fade as the circumbinary nebula enters eclipse by the red giant and then slowly recover at the end of the eclipse. We estimate this duration to be 86.8 days. The two points labelled B at magnitude 12.40 mark the start of ingress and end of egress of eclipse of the extended hot component behind the red giant. This eclipse lasted 55.1 days. Points labelled C at magnitude 13.18 mark the midpoints by magnitude of eclipse ingress and egress. We consider these points as marking the times at which the white dwarf in the centre of the hot component entered and emerged from eclipse by the red giant. The white dwarf eclipse therefore lasted 35.1 days. Points labelled D at magnitude 13.96 mark the start and end of total eclipse of the extended hot component which lasted 14.9 days. These durations may be used to estimate the size of the hot component and the orbital inclination. In doing this we are assuming the giant star is spherical with a well-defined surface. The detrended V magnitudes, orbital phases, and JD corresponding to these points of the eclipse are listed in Table 1. Uncertainty in determining the phase of these points is estimated as 0.002, giving an uncertainty in JD of 0.9 day.

The detrended out-of-eclipse magnitude is 12.11 and the mean V magnitude during the minimum is $14.00 \pm 0.02$. The depth of the eclipse in the V band is therefore 1.89 magnitudes, a reduction in flux of 82.5%. Thus $> 80\%$ of the visual light from the system comes from the eclipsed hot component and its immediate surroundings while $< 20\%$ comes from the red giant and uneclipsed nebular light.

The observed eclipse profile is not consistent with the eclipse of a nominally-sized white dwarf as the ingress and egress stages of the eclipse would have been vertical separated by a longer period of totality. The profile observed suggests that the hot component is an extended luminous object surrounding the white dwarf which is progressively eclipsed, disappears for a period, and then slowly emerges. The nature of this object is debated in the literature as being either an accretion disc or the apparent photosphere of a hot giant star (Esipov *et al.* 2013; Tatarnikova *et al.* 2020), or possibly something else.

## 8. Component parameters

There is little definitive information about the physical parameters of V1413 Aql in the literature. Gałan *et al.* (2023) say in an Appendix that there is no spectroscopic orbit available for V1413 Aql and no certain information available about the component parameters. Neither is there any information on the components or orbital parameters of V1413 Aql in the New Online Database of Symbiotic Variables (Merc *et al.* 2019) or in the "Catalogue of Symbiotic Stars" by Belczyński *et al.* (2000).

We will assume a nominal white dwarf mass of $0.6\,M_\odot$ (Blouin 2024). Tatarnikova *et al.* (2009) adopt this value and a mass of $1.2\,M_\odot$ for the red giant. They also give $130\,R_\odot$ as the radius of the giant star. This is broadly consistent with van Belle *et al.* (1999) who give a statistically fitted radius for an M5III star as $113\,R_\odot$, and van Belle *et al.* (2021) who give a weighted average radius of an M5III star as $98 \pm 30\,R_\odot$. Using the known orbital period and the above masses for the white dwarf and giant components, Kepler's Third Law gives the semi-major axis as $293.5\,R_\odot$ (= 1.36 AU). Based on what they describe as preliminary modelling of photometric data, Siviero *et al.* (2007) give the semi-major axis as 1.7 AU. This value of semi-major axis would imply an unlikely red giant mass of $2.9\,M_\odot$. We will therefore adopt the radius of the giant star as $130\,R_\odot = 0.60$ AU, the semi-major axis of the orbit as 1.36 AU, and assume the orbit is circular.

As noted above, we take 35.1 days as the duration of eclipse of a nominally-sized white dwarf in the center of the hot component by a spherical giant star. In that time the white dwarf will have travelled a distance equal to the width of the giant star as seen from our viewpoint projected onto the orbital plane. This distance depends on the orbital inclination because, as our viewpoint moves up out of the orbital plane, we see a smaller cross section of the giant star projected onto the orbital plane and causing the eclipse. Using the above values for the semi-major axis and radius of the giant, and knowing that the orbital period is 434.1 days, we can find from geometry an estimate of the orbital inclination which gives a duration of 35.1 days for eclipse of the white dwarf by the giant. Allowing for uncertainty in determining the eclipse duration, this orbital inclination is $67.9 \pm 0.3°$. See the Appendix for an explanation of this calculation.

Table 1. Detrended V magnitudes, orbital phases, and Julian Dates corresponding to the key stages of the 2024 eclipse marked in Figure 6.

| Stages of the 2024 eclipse | Detrended V magnitude | Orbital phase | JD |
|---|---|---|---|
| Start of eclipse A | 12.11 | –0.094 | 2460505.0 |
| Start of ingress B | 12.40 | –0.062 | 2460519.0 |
| Midpoint of ingress C | 13.18 | –0.038 | 2460529.5 |
| End of ingress D | 13.96 | –0.019 | 2460537.7 |
| Eclipse minimum | 14.00 | 0.000 | 2460545.8 |
| Start of egress D | 13.96 | 0.016 | 2460552.6 |
| Midpoint of egress C | 13.18 | 0.043 | 2460564.6 |
| End of egress B | 12.40 | 0.065 | 2460574.1 |
| End of eclipse A | 12.11 | 0.106 | 2460591.8 |

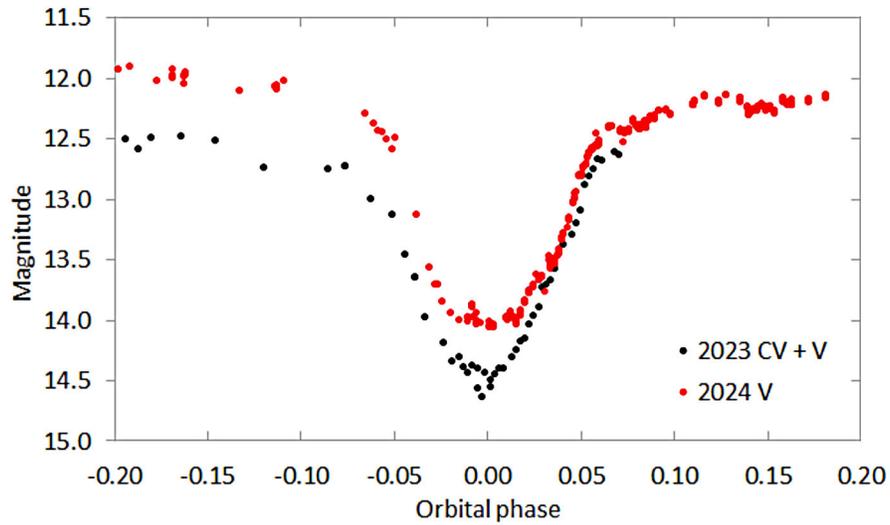

Figure 4. Data from the AAVSO AID for eclipses of V1413 Aql in 2023 and 2024 relative to orbital phase of the 2024 eclipse.

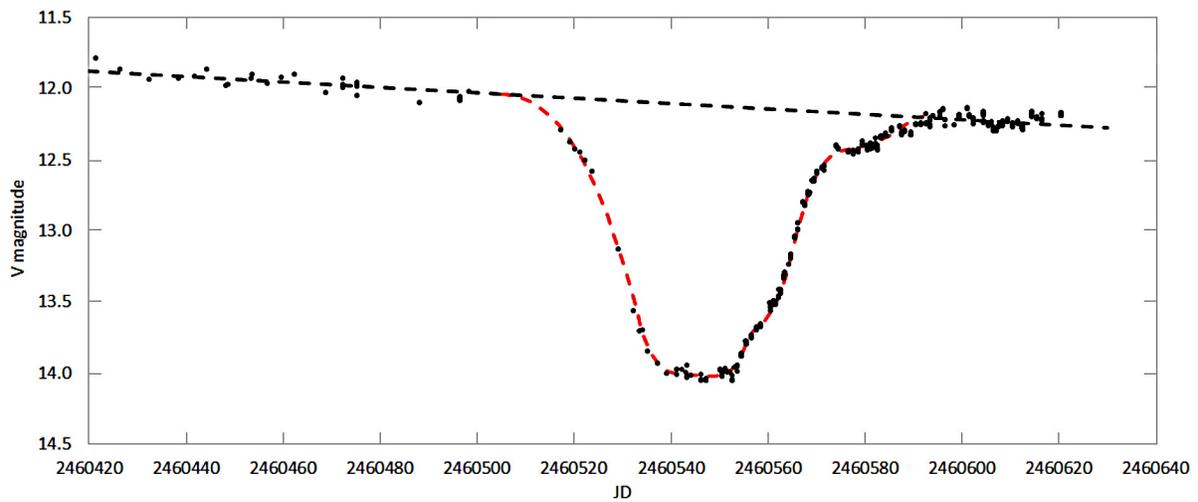

Figure 5. V-band magnitudes of V1413 Aql in the AAVSO AID prior to the eclipse together with our V magnitudes, all in black. The steady downward trend throughout the eclipse is marked as a dashed line in black. The multiple polynomial fits to our V magnitudes are shown as a dashed line in red.

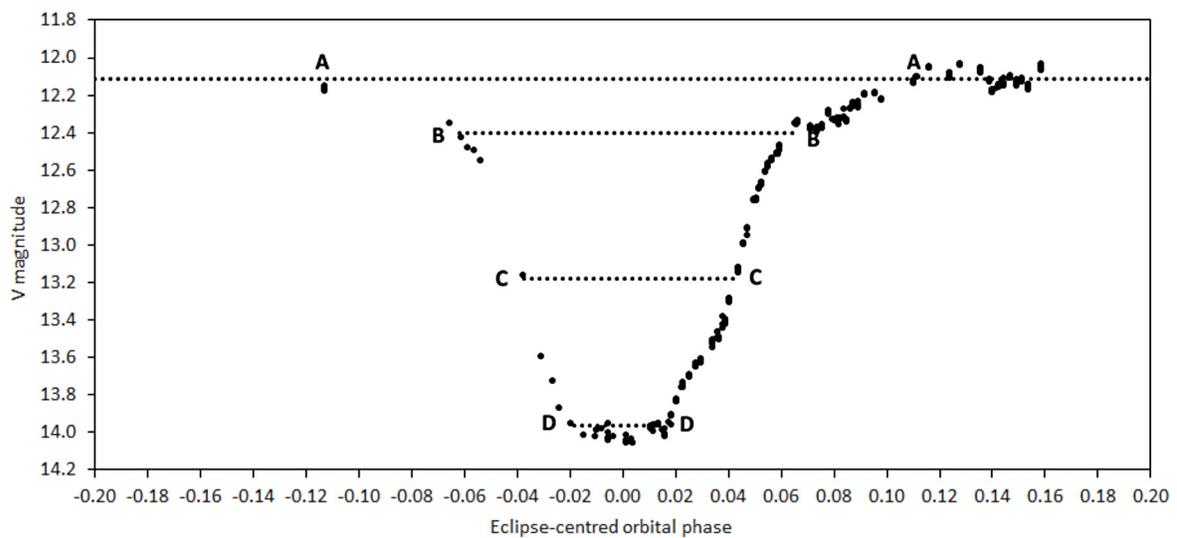

Figure 6. Detrended V magnitudes of V1413 Aql during the 2024 eclipse with magnitudes of key stages of the eclipse marked by horizontal dotted lines (see text).

We are taking duration of the eclipse of the extended hot component by the giant star from start to finish as 55.1 days. Of this, $55.1 - 35.1 = 20.0$ days is the time it took the hot component to travel its own diameter in its orbit. From this, and the above inclination, we can calculate the radius of the extended hot component as $39.3 \pm 1.7 R_\odot = 0.183 \pm 0.008$ AU ($= 30\%$ of the radius of the giant star). This makes the separation between the face of the giant star and the nearest edge of the extended hot component to be $293.5 - 130 - 39.3 = 124.2 R_\odot = 0.5$ AU. The white dwarf and extended hot component are totally eclipsed behind the giant star for $35.1 - 20.0 = 15.1$ days. This is consistent with the observed value of 14.9 days.

Using the formula in Eggleton (1983) to calculate the radii of the Roche lobes (RL) of the red giant and hot component with our adopted masses gives the RL radius as 0.60 AU for the giant and 0.44 AU for the hot component. This suggests the red giant in V1413 Aql is completely filling its RL and the hot component RL is only 41% full by radius. This suggests the giant star may be transferring matter to the white dwarf, possibly via an accretion disc.

## 9. Spectral analysis

Figure 7a to d show examples of our flux calibrated and dereddened spectra recorded at different stages of the eclipse with prominent emission lines identified. These spectra contain contributions from the red giant, the extended hot component, and the nebula in different proportions during the eclipse. Only weak evidence of the TiO molecular absorption bands expected in the spectrum of a M5 giant star can be seen in spectra in the region beyond 6000 Å. This indicates that light from other sources is veiling these absorption bands over most of the spectrum. The emission lines are caused when radiation from the white dwarf ionizes atoms and heats the gas in the wind of the red giant. The H I, He I, and Fe II lines seen in Figure 7 can be produced by ionizing radiation with photon energies < 25 eV. The forbidden lines of [O III] and [N II] are probably produced in the tenuous outer regions of the nebula, perhaps in low density bubbles produced by shocks. The absence of He II 4686 and other high ionization lines such as [Fe VII] and [Ne V] is evidence that V1413 Aql was still in a lower temperature state and had not reached the hot, visually faint state at the time of the eclipse.

The spectrum of V1413 Aql we observe during eclipse minimum is a combination of the red giant spectrum, the nebular continuum, and emission lines produced in the giant wind by radiation from the white dwarf. By scaling the flux in a M5III spectrum from the Fluks Spectral Flux Library (Fluks *et al.* 1994) to match the profile of the V1413 Aql spectrum at minimum in the region above 6000 Å and subtracting this from the spectrum at minimum, we reasonably obtain the spectrum of the nebular continuum plus emission lines. To extract just the spectrum of emission lines, we needed to find a representation for the underlying nebular continuum. As we do not have a physical model for this, we adopted the empirical approach of looking for a representation of the continuum which closely matched the continuum of our spectra over the wavelength range from 4700 to 7000 Å. This covers the transmission of our V band filter and the majority of the emission lines. We eventually decided to use a black body spectrum as this matched the continuum profile well over this range. We found the best match to be with a black body temperature of 6000 K scaled to match the flux of the continuum profile as shown in Figure 8 a. In doing this we do not assign any physical significance to the use of a black body spectrum, simply that it provided a good match to our requirement. We then subtracted this scaled black body spectrum from the nebular continuum plus emission lines to obtain the flux in the nebular emission lines shown in Figure 8 b.

The spectrum of V1413 Aql we observed outside eclipse is the spectrum observed at minimum plus the spectrum of the extended hot component. By subtracting the spectrum at minimum from the spectrum outside eclipse, we obtained the spectrum of the extended hot component plus emission lines. By comparing the extended hot component spectrum with black body spectra, we again found a close match to the continuum over the above wavelength range with a scaled black body temperature of 6000 K as shown Figure 9 a. By subtracting this scaled black body spectrum from the continuum of the extended hot component plus emission lines, we obtained the flux in the emission lines of the extended hot component shown in Figure 9 b.

In Table 2 we compare the dereddened flux emitted in the V band, equivalent V magnitudes and the proportion of total flux produced by the red giant, the continuum and emission lines of the extended hot component and the continuum and emission lines of the circumbinary nebula. The brightest component is the continuum and emission lines of the extended hot component, which are together responsible for 83% of the light we observe in the V band. The nebula and its emission lines contribute over 14% while the red giant is responsible for less than 3%.

## 10. Emission line flux

As our spectra are calibrated in absolute flux and have been dereddened, we can measure the dereddened flux produced in each emission line although our limited spectral resolution causes some close lines to be blended. To obtain the flux in an emission line we measure the average flux in the continuum on both sides of the line, linearly interpolate the continuum under the line, and subtract this from the spectral profile of the line. This gives us the line profile above a local continuum of zero. We then fit a Gaussian to this profile and integrate the flux under the fit. We found using a Gaussian fit better than numerically integrating the profile of the line because, at the low resolution of our spectra, many emission lines are blended with adjacent lines whose flux may be included in the numerical integration, thus inflating the line flux. Gaussian fits were better at excluding this unwanted flux. In the case of very close lines such as [O III] 5007 and He I 5016, we fitted Gaussians to both lines simultaneously. Uncertainty in line flux is estimated as a combination of the uncertainties in determining the spectral profile in the region of the line and in estimating the interpolated continuum under the line.

Table 2. Dereddened flux emitted in the V band in our spectra, equivalent V magnitude, and proportion of total V band flux for each component of V1413 Aql.

| Component | V band flux (erg/cm2/s) | Equivalent V magnitude | Proportion of Total V band flux (%) |
|---|---|---|---|
| Red giant | 1.36E–12 | 16.04 | 2.71 |
| Extended hot component continuum | 4.04E–11 | 12.35 | 80.44 |
| Extended hot component emission lines | 1.24E–12 | 16.13 | 2.48 |
| Nebular continuum | 5.61E–12 | 14.50 | 11.17 |
| Nebular emission lines | 1.61E–12 | 15.85 | 3.21 |
| Total | 5.02E–11 | 12.12 | 100.00 |

Table 3. Dereddened flux in emission lines in V1413 Aql produced in the nebula and in the extended hot component with their sum and ratio.

| Emission line | Flux in nebula (erg/cm2/s) | Flux in extended hot component (erg/cm2/s) | Sum of fluxes in nebula and hot component | Flux ratio nebula / hot component |
|---|---|---|---|---|
| Hα | 4.95E–12 | 2.90E–12 | 7.85E–12 | 1.71 |
| Hβ | 1.17E–12 | 7.85E–13 | 1.96E–12 | 1.49 |
| Hγ | 4.47E–13 | 4.93E–13 | 9.40E–13 | 0.91 |
| He I 5876 | 2.52E–13 | 2.50E–13 | 5.02E–13 | 1.01 |
| He I 6678 | 6.88E–14 | 6.91E–14 | 1.38E–13 | 0.99 |
| He I 7065 | 9.73E–14 | 1.03E–13 | 2.01E–13 | 0.94 |
| Fe II 4924 | 8.44E–14 | 4.82E–14 | 1.33E–13 | 1.75 |
| Fe II 5198 | 2.91E–14 | 4.22E–14 | 7.13E–14 | 0.69 |
| Fe II 5317 | 7.73E–14 | 2.22E–14 | 9.95E–14 | 3.49 |
| Fe II 6384 | 1.43E–14 | 3.23E–14 | 4.66E–14 | 0.44 |
| [O III] 4959 | 9.08E–14 | — | — | — |
| [O III] 5007 | 2.52E–13 | — | — | — |
| [N II] 5755 | 2.11E-14 | — | — | — |

Figures 10 to 13 show how dereddened emission line flux varied with orbital phase for those lines which could be measured in our spectra. The lines all show a reduction in flux during the eclipse but with different behavior with orbital phase. Variation of the H I Balmer lines is clearly asymmetric in phase. The H I Balmer and some He I and Fe II lines show the largest reductions in flux with drops of 50 to 60%. In most cases the flux level after the eclipse returns to pre-eclipse levels but flux in the [O III] 4959 line decreases steadily through the eclipse as, to a lesser extent, does flux in the [O III] 5007 line. This may point to a cause other than the eclipse for the changes in those lines. This information should be useful for future modelling of the components and physical conditions of the system.

We measure the mean Hα/Hβ, He I 6678/5876, and He I 7065/5876 flux ratios in our data as 4.34 ± 0.28, 0.32 ± 0.04 and 0.28 ± 0.05, respectively. None of these ratios, which are temperature sensitive, show any significant variation during the eclipse. Proga *et al.* (1994) gave the He I 6678/5876 ratio for AS 338 (V1413 Aql) as 0.39 with an uncertainty of ± 15%. Their measurements were uncorrected for reddening and were made at orbital quadrature and therefore outside eclipse. Tatarnikova *et al.* (2020) reported line fluxes during an eclipse in 2012 while V1413 Aql was in an active state and at a similar magnitude to this eclipse. The mean of their He I 6678/5876 ratios during the eclipse was 0.45 ± 0.04. Their H I and He I lines showed similar behavior to ours with flux reducing during the eclipse. Munari (1992) also gave He I flux values measured in plate scans for AS 338 which gave an average value for the He I 6678/5876 ratio of 0.43 ± 0.06, but again these were not recorded during an eclipse. Munari also reported reductions in the flux of H I, He I, and [O III] lines during eclipse broadly consistent with our results.

As well as measuring how emission in each line varied during the eclipse, we also measured how much flux was produced in each line in the nebula and in the extended hot component and their ratio. In Table 3 we compare the dereddened flux measured by Gaussian fits to H I, He I, Fe II, [O III], and [N II] emission lines produced in the nebula (Figure 8 b) and in the extended hot component (Figure 9 b). Flux in the [O III] and [N II] lines could only be reliably measured for nebular lines as these form in the tenuous outer region of the nebula. The sodium D lines were not measurable at our resolution because of their proximity to the strong He I 5876 line, but were visible in emission during the eclipse minimum and in absorption during the early and late stages of the eclipse (Figure 7).

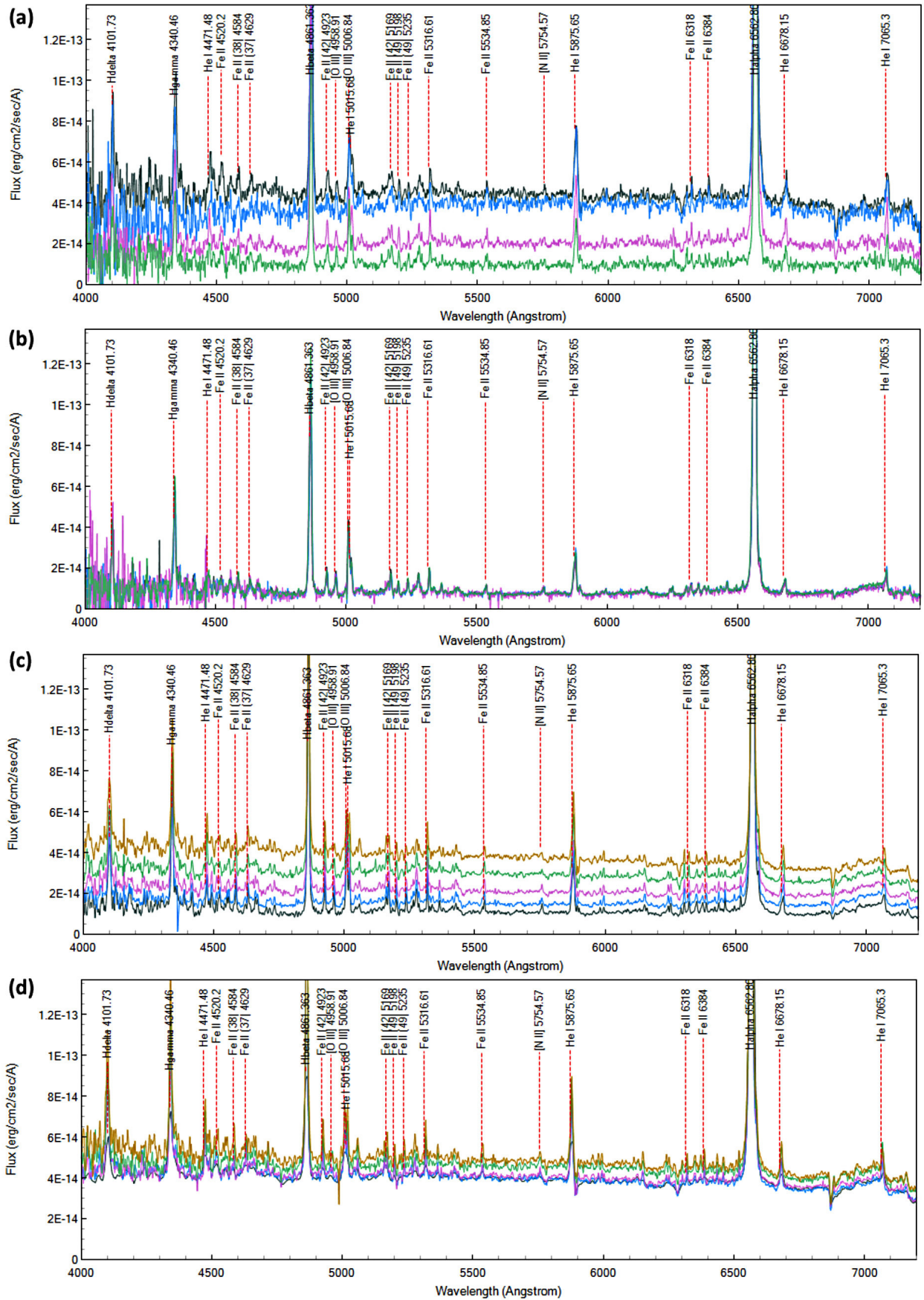

Figure 7. Examples of flux calibrated and dereddened spectra of V1413 Aql recorded during (a) ingress, (b) minimum, (c) egress, (d) end of eclipse, with prominent emission lines identified. All are shown at the same flux scale.

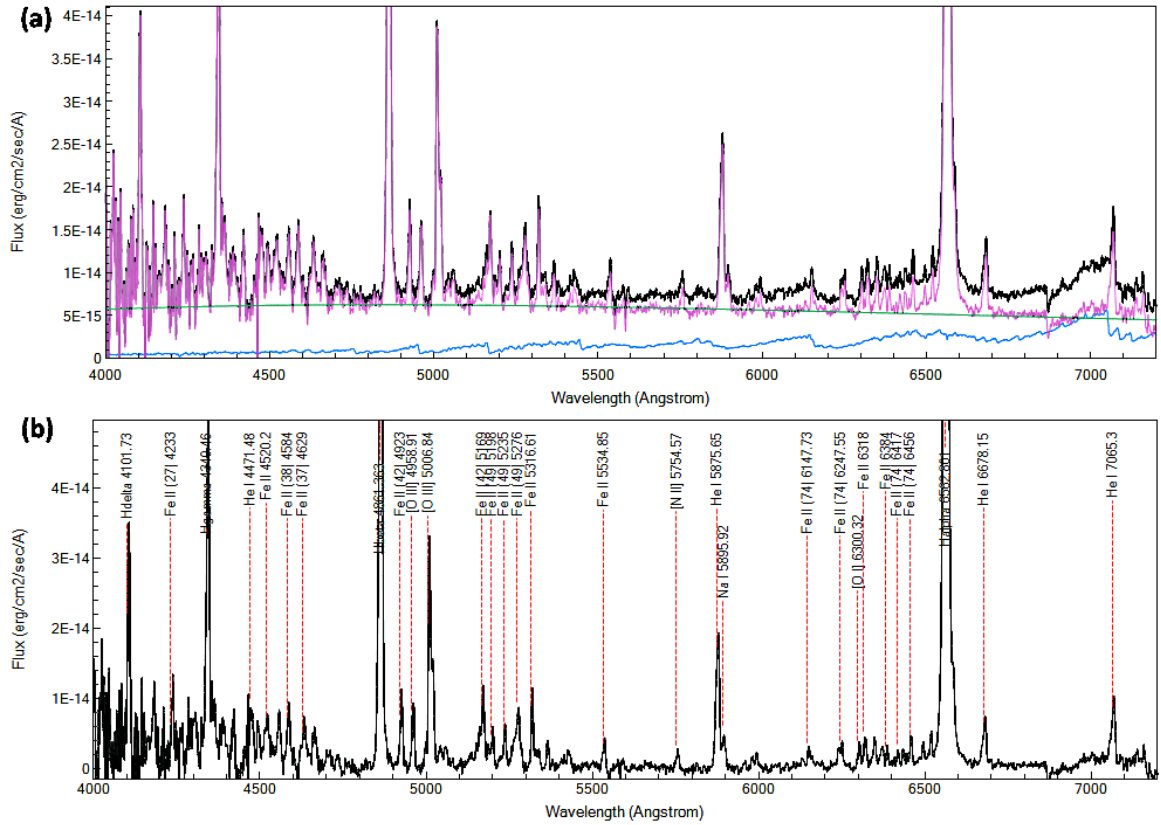

Figure 8. (a) Dereddened spectrum of V1413 Aql at minimum (black), scaled M5III spectrum (blue), nebular continuum plus emission lines (pink) and a scaled black body spectrum (green). (b) Emission lines above the nebular continuum after subtracting the scaled black body spectrum.

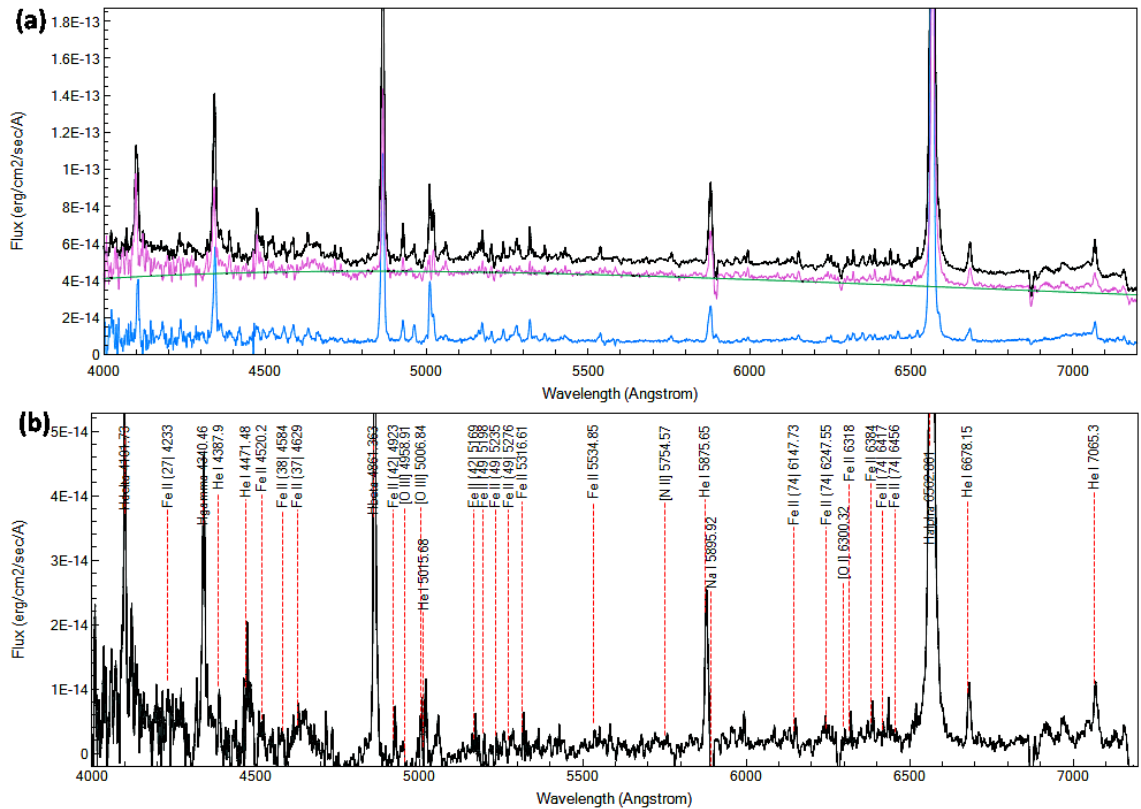

Figure 9. (a) Dereddened spectrum of V1413 Aql outside eclipse (black), spectrum at minimum (blue), difference is equal to the spectrum of the extended hot component (pink) and a scaled black body spectrum (green). (b) Emission lines in the spectrum of the extended hot component after subtracting the scaled black body spectrum.

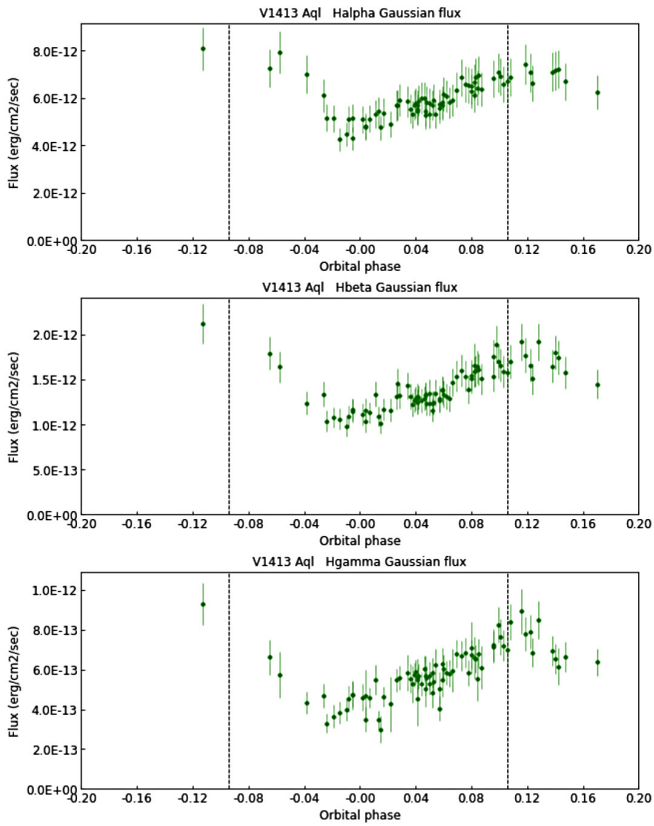

Figure 10. Variation of dereddened H I Balmer emission line flux with orbital phase. Start and end of the eclipse are marked with vertical lines.

## 11. Summary

We have used standard filtered photometry and low resolution flux calibrated and dereddened spectroscopy recorded during the 2024 eclipse of the symbiotic binary V1413 Aql to investigate the components of the system and the nature of its emission spectrum. The value of the B–V color index and absence of high ionization emission lines indicated the system was still in an active state at the time of the eclipse. We looked for, but did not see, flickering as evidence of accretion. This eclipse was 4.6 days late relative to the Munari (1992) ephemeris but within its published uncertainty. The eclipse profile showed that the hot component was an extended object rather than an isolated white dwarf. By analyzing the eclipse profile and making reasonable assumptions about the masses of the components and the radius of the red giant, we determined the orbital inclination to be $67.9 \pm 0.3°$, the radius of the extended hot component surrounding the white dwarf to be $39.3 \pm 1.7 R_\odot$, and that the red giant star was probably filling its Roche Lobe. Future spectroscopic observations around the binary orbit should enable more precise determination of the orbital parameters. By analyzing our flux calibrated spectra, we determined that the brightest component of the system is the extended hot component whose continuum and emission lines are together responsible for 83% of the light we observe in the V band. The circumbinary nebula and its emission lines contribute over 14% while the red giant is responsible for less than 3%. Our spectra revealed a rich harvest of low ionization emission lines. We measured the flux of each line in the nebula and in the extended hot component and their ratio. By measuring how flux in these lines varied through the eclipse, we have provided information which should prove useful for future modelling of the components and physical conditions of this symbiotic system.

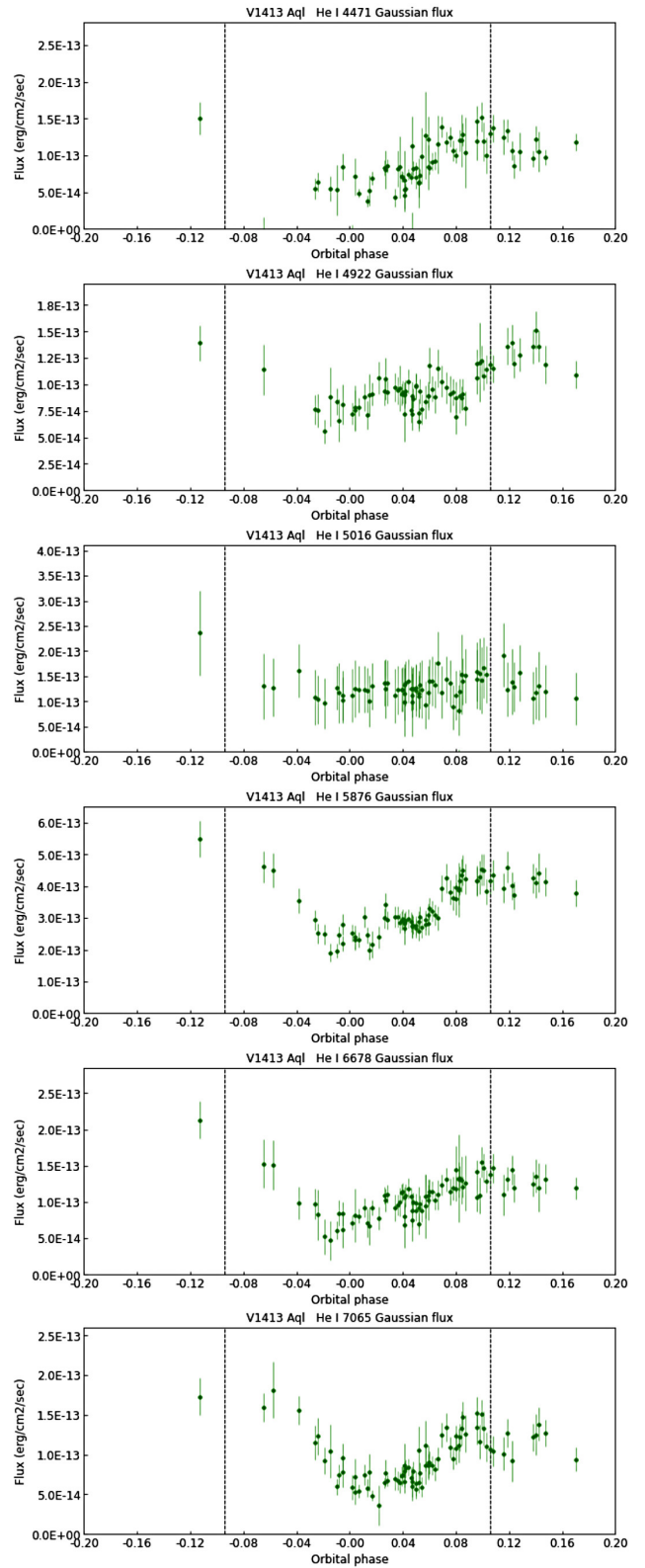

Figure 11. Variation of dereddened He I emission line flux with orbital phase. Start and end of the eclipse are marked with vertical lines.

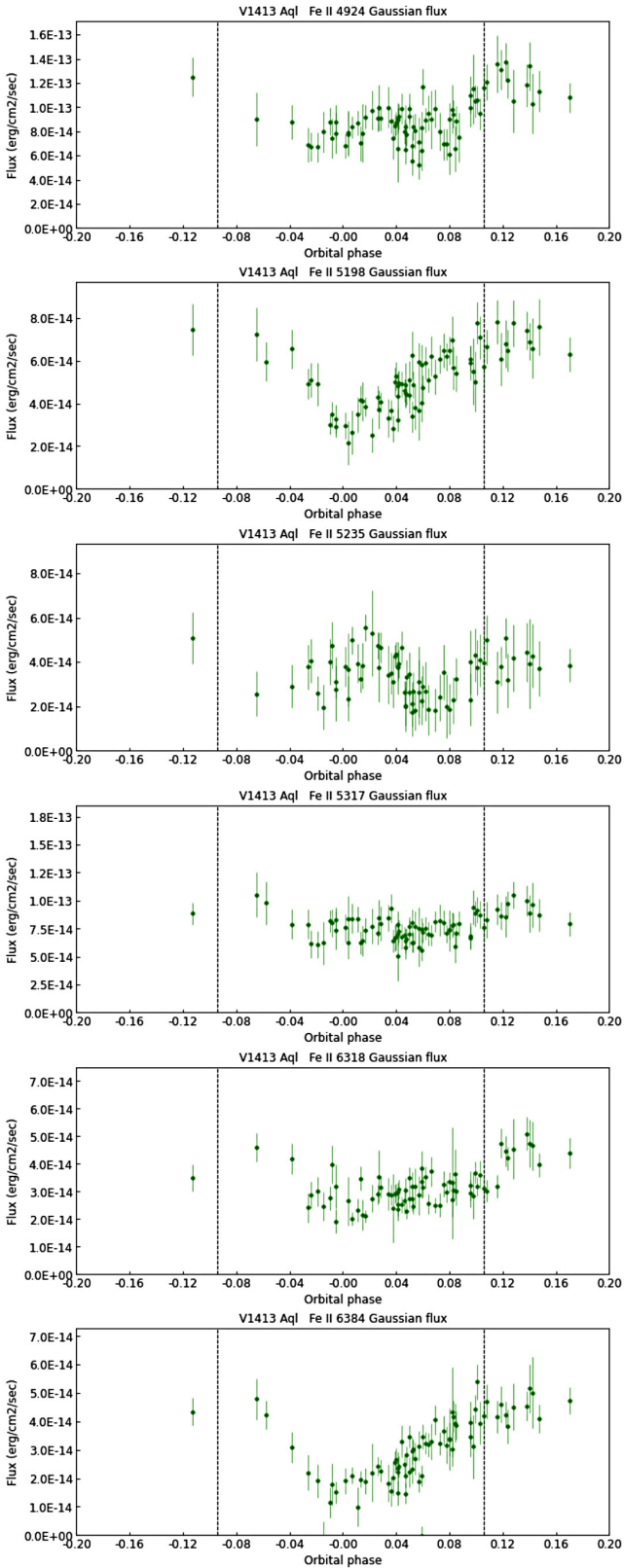

Figure 12. Variation of dereddened Fe II emission line flux with orbital phase. Start and end of the eclipse are marked with vertical lines.

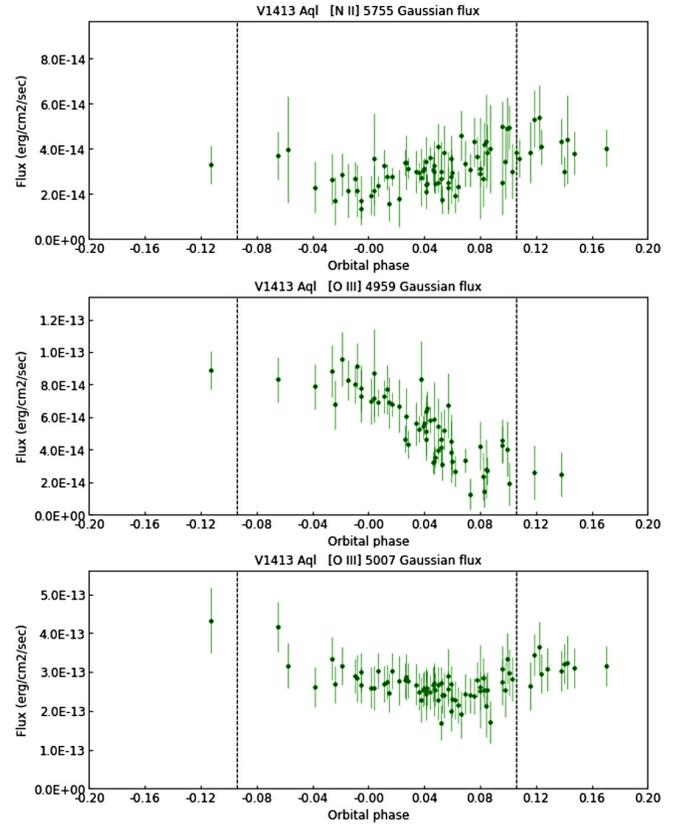

Figure 13. Variation of dereddened [N II] and [O III] emission line flux with orbital phase. Start and end of the eclipse are marked with vertical lines.

## 12. Acknowledgements

We thank the anonymous referee for comments which have helped us to improve the paper. We are grateful to Prof. Dr. Hans Martin Schmid and Dr. Chris Lloyd for helpful comments and suggestions. Francois Teyssier set up and continues to manage the ARAS Spectral Database. The BAA Photometry and Spectroscopy Databases were developed and are maintained by Dr. Andy Wilson. We acknowledge with thanks the variable star observations from the AAVSO International Database contributed by observers worldwide and used in this research. We also acknowledge the work of developers of the Astropy package and of Mr. Tim Lester, developer of PLOTSPECTRA.

## Appendix: calculation of orbital inclination

This calculation of orbital inclination is a simple geometrical analysis consistent with our observational data rather than a definitive physical analysis of the parameters of V1413 Aql.

P = orbital period = 434.1 days.
a = assumed semi-major axis of the circular orbit = 1.36 AU.
r = assumed radius of the red giant = 130 $R_\odot$.
T = observed duration of white dwarf eclipse = 35.1 days.
i = unknown orbital inclination.

The line of sight from the observer to the white dwarf passes a distance c above the center of the red giant where c = a*cos(i).

This line of sight cuts through the red giant in a circular plane whose radius d = $\sqrt{(r^2 - c^2)}$.

It is the part of the red giant defined by this circular plane which eclipses the white dwarf.

The distance from the center of this circular plane to the white dwarf is b = a*sin(i).

The circular plane projects a shadow of width 2d onto the orbit of the white dwarf.

This part of the white dwarf orbit in the shadow of the red giant subtends an angle 2e at the center of the above circular plane where e = $\sin^{-1}(d/b)$.

This represents a fraction f of the white dwarf orbit where f = 2e/360.

This is the same fraction as the ratio of the white dwarf eclipse to the orbital period, so f = T/P.

The value of i which equates these fractions is 67.9°.

The uncertainty in i is derived from the uncertainty in measuring the duration of the white dwarf eclipse.

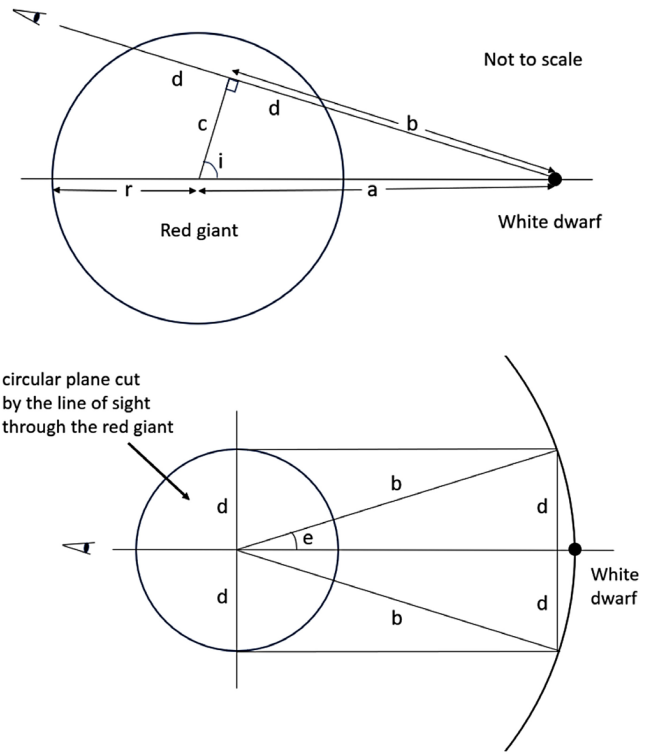

Figure A1. A simple geometrical analysis of V1413 Aql.